# Streamlined Inexpensive Integration of a Growth Facility and Scanning Tunneling Microscope for *in situ* Characterization

P. Xu, D. Qi, S.D. Barber, C.T. Cook, M.L. Ackerman, and P.M. Thibado[a]

*Department of Physics, University of Arkansas, Fayetteville, Arkansas 72701*

**Abstract**

The integration of a scanning tunneling microscope chamber with a sample growth facility using non-custom, commercially available parts is described. The facility also features a newly-designed magnetic wobble stick to increase the reliability of sample transfer in a cost-effective manner.

The limitations of silicon-based technology have become increasingly evident, leading research efforts to focus on more promising materials, such as graphene. In contrast to silicon, graphene's most important properties reside on its surface, rather than on the characteristics of its bulk. Surface techniques, such as scanning tunneling microscopy (STM), are therefore required to manufacture and study graphene.

Graphene is manufactured using a variety of methods, such as mechanical exfoliation of graphite, thermal reduction of silicon carbide, and precipitation of carbon dissolved in transition metals.[1] Most recently, the production method of choice has been epitaxial growth of graphene on a metal substrate (such as nickel or copper) using chemical vapor deposition (CVD),[2] as well as a related technique which involves the deposition of pure carbon in a vacuum environment on a heated copper substrate. While the latter method is becoming increasingly popular for producing graphene for research purposes, much remains to be learned about the procedure. For example, metal substrates typically exhibit widely varied surface topography, and it is yet to be explained how this affects the nucleation and growth of graphene. Another area in need of clarification is the nucleation process and its role in determining the orientation of the graphene. Finally, more can be learned about the impact a step edge has on the graphene as it grows across the surface from a nucleation site, as well as what occurs when growths from two distinct nucleation sites meet. Examining the surface of samples with STM at various points throughout the growth process may yield insight to these questions.

---

[a] Electronic mail: thibado@uark.edu

For such STM studies, several conditions must be met. First of all, in most cases, the substrate must be heated to 1000 °C to remove the surface oxide in preparation for growth. Secondly, the sample must be transferred in vacuum between facilities – a growth facility (GF) and an STM - to avoid contamination of its surface. Finally, the sample must be held rigidly to the STM stage to ensure vibration isolation. These conditions can be met by connecting the STM chamber to the GF so that samples may be transferred between the two in a controlled environment. Such a setup allows monitoring of the various stages of the graphene growth process by periodically halting growth and then imaging the surface.

Currently, only a few such combined STM/GF systems[3-4] have successfully performed studies related to a fundamental understanding of materials growth processes,[5] leading to numerous and significant discoveries. Most of these systems incorporate customized components, either in the STM imaging stage, the growth facility, or within the sample handling in the transfer sections between them. In this note, we detail a novel integration approach which requires no customization. A commercially available ultrahigh vacuum (UHV) STM (Omicron) and GF (Riber) were the starting point. The design strategy, however, is compatible with any STM and growth facility. Its major advantage is that the GF and STM themselves were not customized, nor were the tools for sample transport. In addition, both systems can continue to work independently without any interaction or contamination. We also developed, in collaboration with a company,[6] a magnetic wobble stick. It brought advantages to the system as well, as it is more robust than its traditional counterpart and does not require a costly edge-welded bellows.

From an overall perspective, this approach connected the GF to the STM facility using all stainless steel, UHV components. A top view of the GF-STM multi-component facility is shown in Fig. 1. The GF chamber is represented by a large square in the upper middle of the diagram, and is attached to the GF transfer chamber shown directly below it. This chamber is simply a 6-way cross chamber with 8-inch ports on each side. The magnetic GF transfer arm extends below the GF transfer chamber in the diagram. It transports samples between the GF chamber and the GF transfer chamber. The STM chamber is represented by a large circle on the right, attaching to the STM transfer chamber shown directly below it. Extending below the STM transfer chamber is the STM magnetic transfer arm, which transfers samples between the STM chamber and the STM transfer chamber.

The GF and STM transfer chambers were connected using a third interconnection chamber, a setup which requires fewer components and transfers than alternative techniques, such as rolling a sample cart between the chambers.[7] The sample is moved through the interconnection chamber using the GF-STM magnetic transfer arm, which enters the GF transfer chamber from the left in Fig. 1.

In order to more clearly illustrate the relationship between the transfer points, a front view of the GF-STM multi-component facility is provided in Fig. 2. The GF and STM magnetic transfer arms are coming out of the page in this view. The newly-designed magnetic wobble stick is visible, extending from above into the GF transfer chamber at an angle of about 30° off of the vertical direction. The additional ports positioned on top of the GF transfer chamber allow viewing and lighting. To transport a sample, the magnetic wobble stick grasps it and moves it between the GF transfer arm and the GF-STM transfer arm. The GF-STM transfer arm then moves it to the STM transfer chamber, where the STM transfer arm accepts it. Once inside the STM transfer chamber, the sample could be moved into the STM chamber or removed from the vacuum chamber entirely via a 4 ½-inch sample load lock view port. Correspondingly, new samples can enter the vacuum system from the ambient conditions of the room through this port.

The interconnection chamber between the GF and the STM has several components which are critical to successful integration. A front view of this chamber is shown in Fig. 3. On the left end, an all-metal valve facilitates venting and pumping. To the right of that, a gate valve is shown which allows venting of one side of the chamber while maintaining vacuum in the other. An ion gauge is positioned near the middle of the chamber to monitor pressure. Because the STM is sensitive to mechanical noise, to the right is an edge-welded bellows to reduce vibrational coupling. If desired, the bellows may be removed to separate the GF and STM chambers from one another. The bellows offers the additional advantage of easy alignment of one transfer chamber with the other, as the interconnection chamber is rigid otherwise. On the far right end is a ceramic section which serves as an electrical break between the GF and STM chambers, which isolates the STM from electrical noise.

A detailed illustration of the newly-designed wobble stick is shown in Fig. 4. The shaded rectangles at the top represent two magnets which are contained in an outer housing (not shown). The lower magnet is magnetically coupled to a hollow tube inside the wobble stick. This tube moves vertically and rotates with the outer casing, much like a typical magnetic drive, offering 360° of continuous rotation and decoupled simultaneous translation along its axis of 20+ inches. This high degree of functionality is not possible with a typical bellows-design wobble stick. The upper magnet is coupled to a stainless steel wire which runs through the center of the inner tube, connecting to one side of the pincer. This magnet simply moves in tandem with the lower magnet unless the two push levers are compressed on the casing. These levers compress a spring between the two magnets, pushing the upper magnet closer to the lower one and causing the wire to open the pincers [see Fig. 4(b)]. When the push levers are released, the compressed spring expands, pulling the wire and closing the pincers. The wobble stick attaches to the GF transfer chamber with a typical 2 ¾-inch Conflat flange, with formed bellows providing angular degrees of freedom of motion (±20°).

Sample handling from the GF transfer chamber to the STM transfer chamber is achieved using standard parts obtained from various system manufacturers, and is illustrated in Fig. 5. The standard STM sample plate is held to a standard GF moly block about two inches in diameter and featuring six pegs around its circumference, as shown in Fig. 5(a) just after growth. Three of the pegs are used to secure the moly block on the end of the GF transfer arm, while the other three serve to hold it in place when it is in the GF chamber. After the moly block is taken from the GF chamber and placed in the GF transfer chamber, the wobble stick is used to remove the STM sample plate from its holder on the moly block. The pincers open and close around the sample plate's eyehole to securely pull it from the pressure clips, as shown in Fig. 5(b). As shown, removing the STM sample plate leaves behind the STM sample plate mount. This mount is the standard mount for the sample plate. Next, the GF transfer arm is pulled out of the way with the empty moly block still attached, while the GF-STM transfer arm is pushed forward into position beneath the wobble stick. The user rotates the wobble stick by 90° and lowers the sample into the front slot on the standard STM sample transfer tool attached to the end of the GF-STM transfer arm, as shown in Fig. 5(c). Finally, the GF-STM transfer arm is fully extended through the interconnection chamber into the STM transfer chamber and rotated by 90°, where the sample can be accepted by the STM transfer arm, as shown in Fig. 5(d). Notice the long rods on the STM transfer arm which enable a mechanism to grasp the eye hole. In summary, throughout the sample transfer and the sample handling parts were commercially obtained.

As evidence of the system's success, room-temperature STM images of graphene grown on copper foil using CVD[8] are shown in Fig. 6. Graphene covers the entire surface in a closely woven fashion. A large scale 1 $\mu$m × 1 $\mu$m image illustrates the surface morphology of graphene on a copper foil substrate as shown in Fig. 6(a). Triangularly- shaped structures dominate the surface at this scale. This view also shows a less common staircase-like structure, with each step spanning about 10 nm and rising about 1 nm. A higher magnification image (200 nm × 200 nm) is shown in Fig. 6(b), featuring a Moiré pattern, which may be the result of two overlapping graphene layers. At the atomic-scale, the honeycomb structure of graphene characterizes the empty-state image shown in Fig. 6(c). A line profile taken from the image shown in Fig. 6(b) is shown in Fig. 6(d). This line profile shows the large height changes that occur across the Moiré pattern where the arrow is drawn.

In summary, we have introduced a convenient and inexpensive method to combine sample growth and STM in one facility. The original sample transport systems have been maintained, with the addition of a magnetic transfer arm and an interconnection chamber. A newly-designed magnetic wobble stick offers continuous rotation, translation, and angular pitch in a low-cost robust design. Overall, this non-custom integration design provides real opportunity for broader use of combined GF-STM systems.

The authors would like to thank Transfer Engineering and Mfg. for their cooperation and assistance in designing the magnetic wobble stick. This work is supported in part by the National Science Foundation (NSF) under grant number DMR-0855358 and the Office of Naval Research (ONR) under grant number N00014-10-1-0181.

Figure Captions

Fig. 1. Top view of the entire multi-chamber facility. The large center square represents the growth facility (GF), while the large circle represents the scanning tunneling microscope (STM). The interconnection chamber bridges the two chambers, and three magnetic transfer arms move the sample from one location to another.

Fig. 2. Front view of the GF-STM multi-chamber UHV facility. This provides a clear view of the magnetic wobble stick and its orientation. It also shows how the GF-STM transfer arm moves the sample between the two transfer chambers.

Fig. 3. A front view and close-up view of the chamber connecting the GF and STM transfer chambers. This chamber contains numerous vital components for successful integration with the STM. The bellows and electrical break are needed to mechanically and electrically isolate the STM from the rest of the system. Also necessary are the vent valve, gate valve, and pressure gauge.

Fig. 4. Schematic illustration of a newly-designed magnetic wobble stick is shown. The wobble stick has numerous enhancements over a traditional wobble stick, including 360° continuous rotation, over 20 inches of linear translation, ±20° of conical angular pitch, and a more robust formed bellows. (a) Shows the pincer in the closed position. (b) Shows the pincer in the open position due to the manually activated levers around the magnets being compressed.

Fig. 5.  Transferring an STM sample plate using the magnetic wobble stick and the STM sample transfer tool is illustrated.  (a) Shows the wobble stick pincer open and approaching the STM sample plate just after growth.  (b) Shows the wobble stick closed and lifting the STM sample plate away from the moly block and STM sample plate mount.  (c) Shows the wobble stick pincer rotated 90° and placing the STM sample plate in the front position of the STM transfer tool. (d) Shows the STM transfer tool after moving from the GF transfer chamber to the STM transfer chamber.  It has also been rotated 90°, then pressed into the two long support rods shown.  At this stage the standard Omicron transfer takes place to grasp the sample and move it into the STM chamber.

Fig. 6.  Room temperature STM images of graphene on copper foil (a) Empty-state (0.3 V), 1 µm × 1 µm STM image showing a staircase structure adjacent to the typical triangularly-shaped structure seen on copper foil.  Each step in the long staircase spans about 10 nm and rises about 1 nm.  (b) Shows a medium scale STM image of bi-layer graphene.  (c) Atomic-scale image showing the honeycomb structure of graphene.  Graphene is found to cover the entire surface in a closely woven fashion.  (d) Line profile taken from (b) showing the height and length scale for the Moiré pattern's features at the location of the drawn arrow.

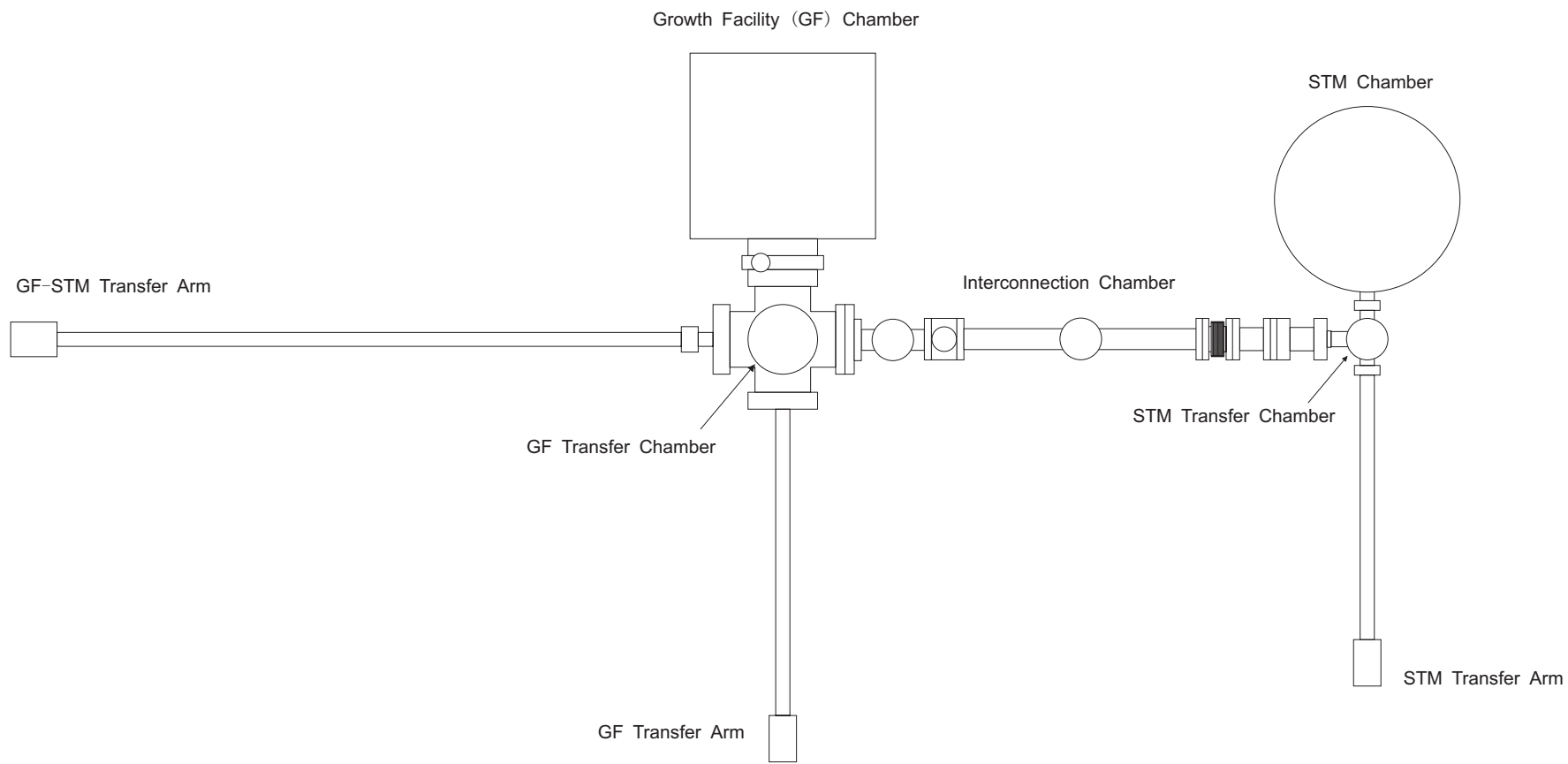

Fig 1. by Xu *et al.*

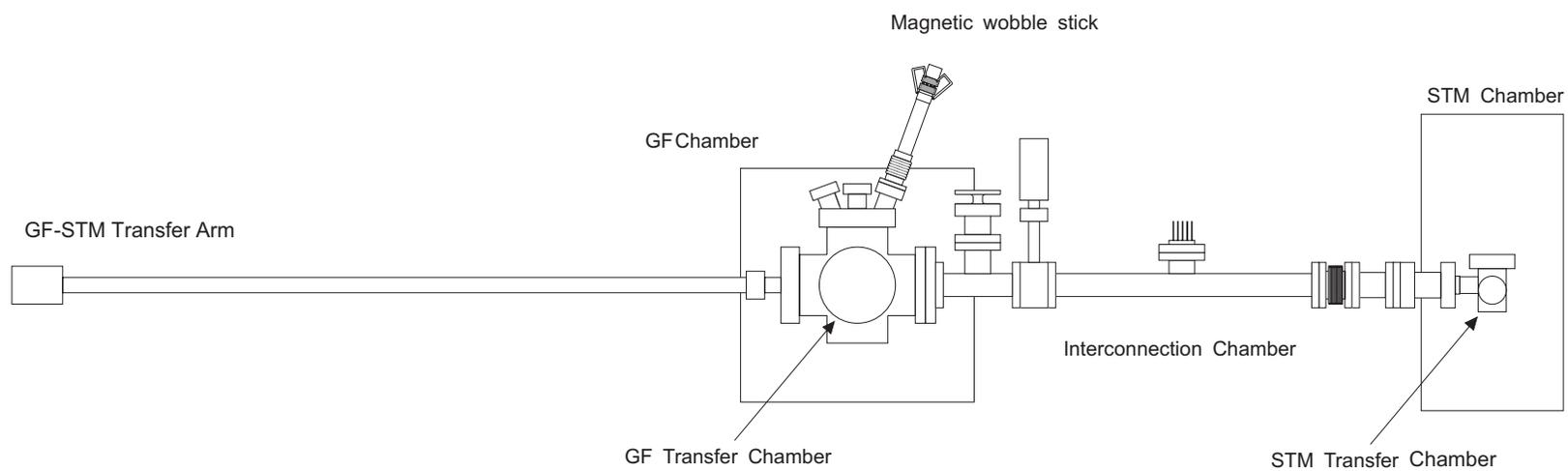

Fig 2. by Xu *et al.*

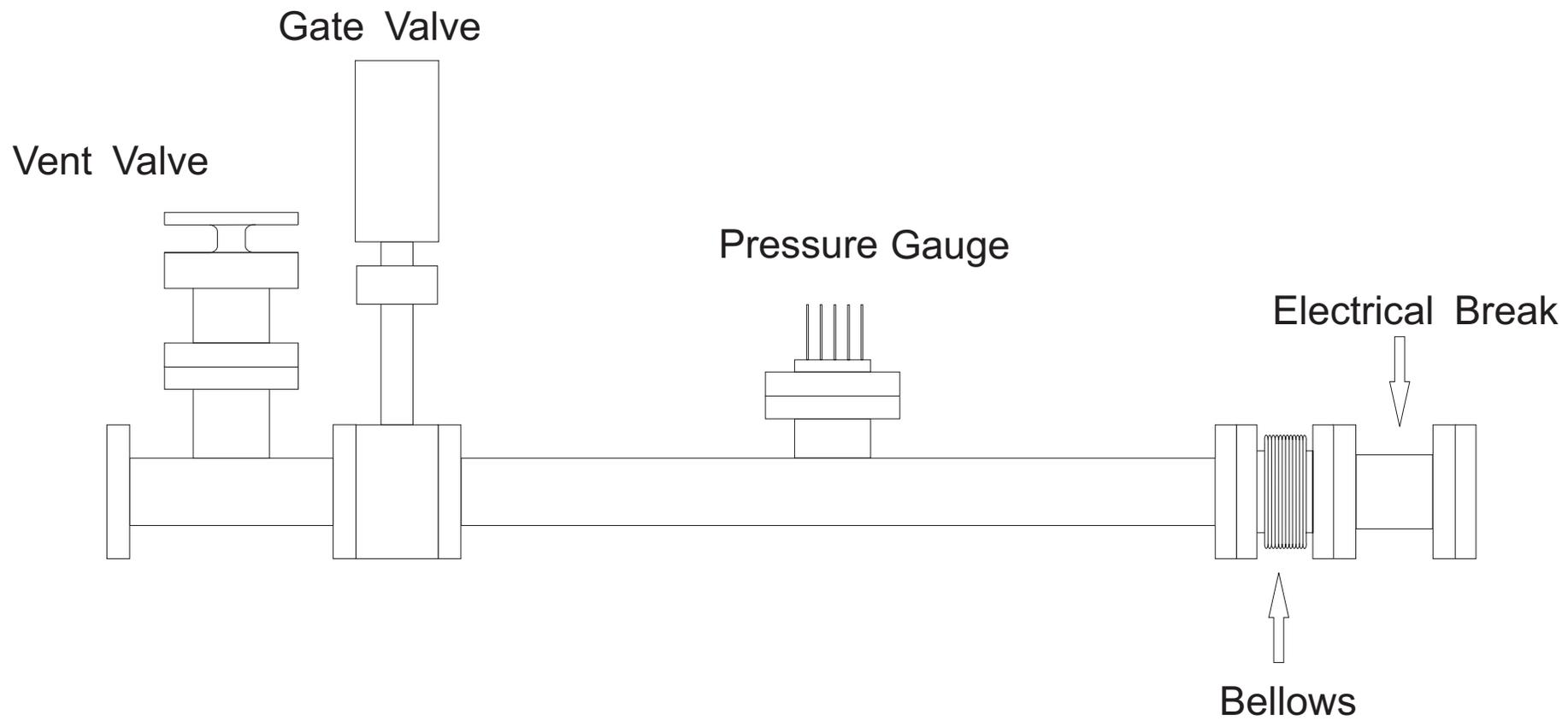

Fig 3. by Xu *et al.*

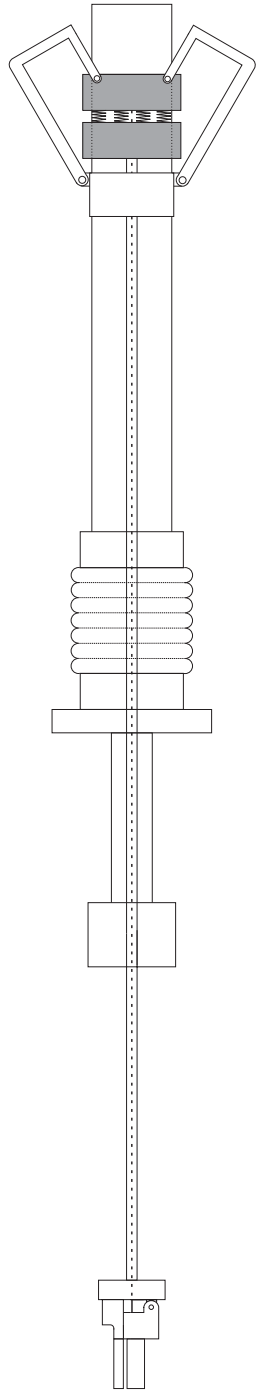 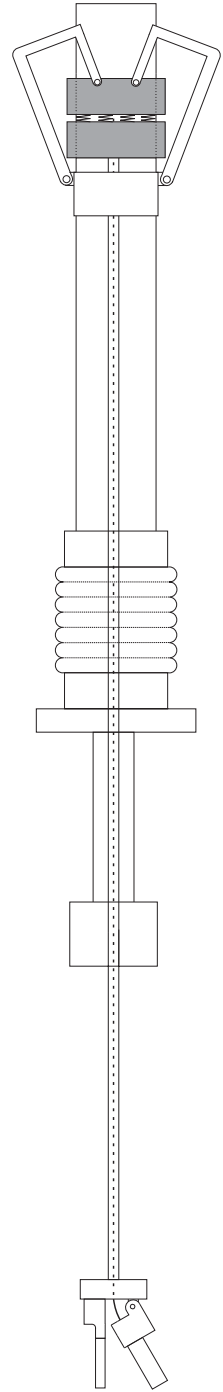

(a) (b)

Fig 4. by Xu *et al.*

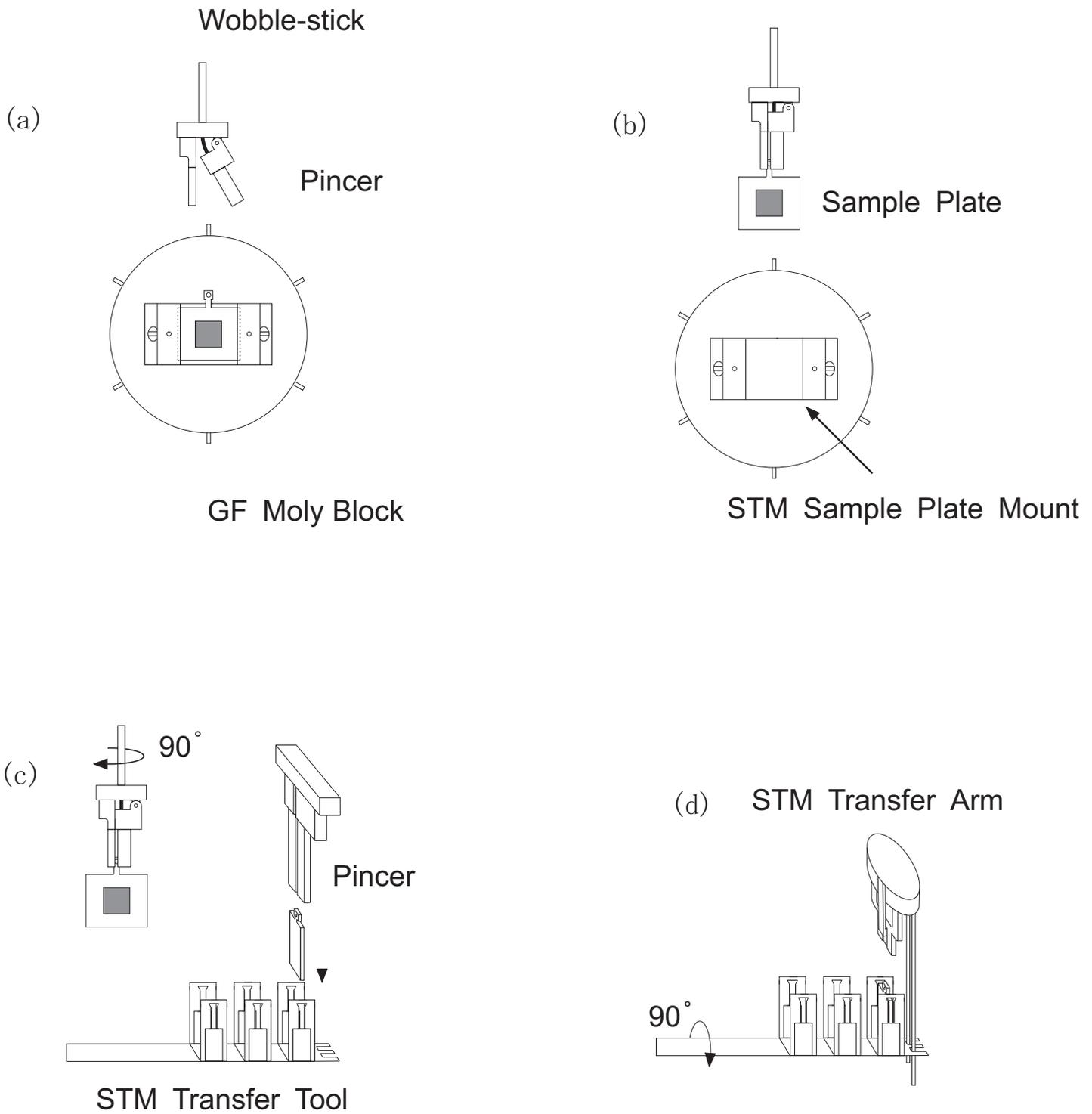

Fig 5. by Xu *et al*.

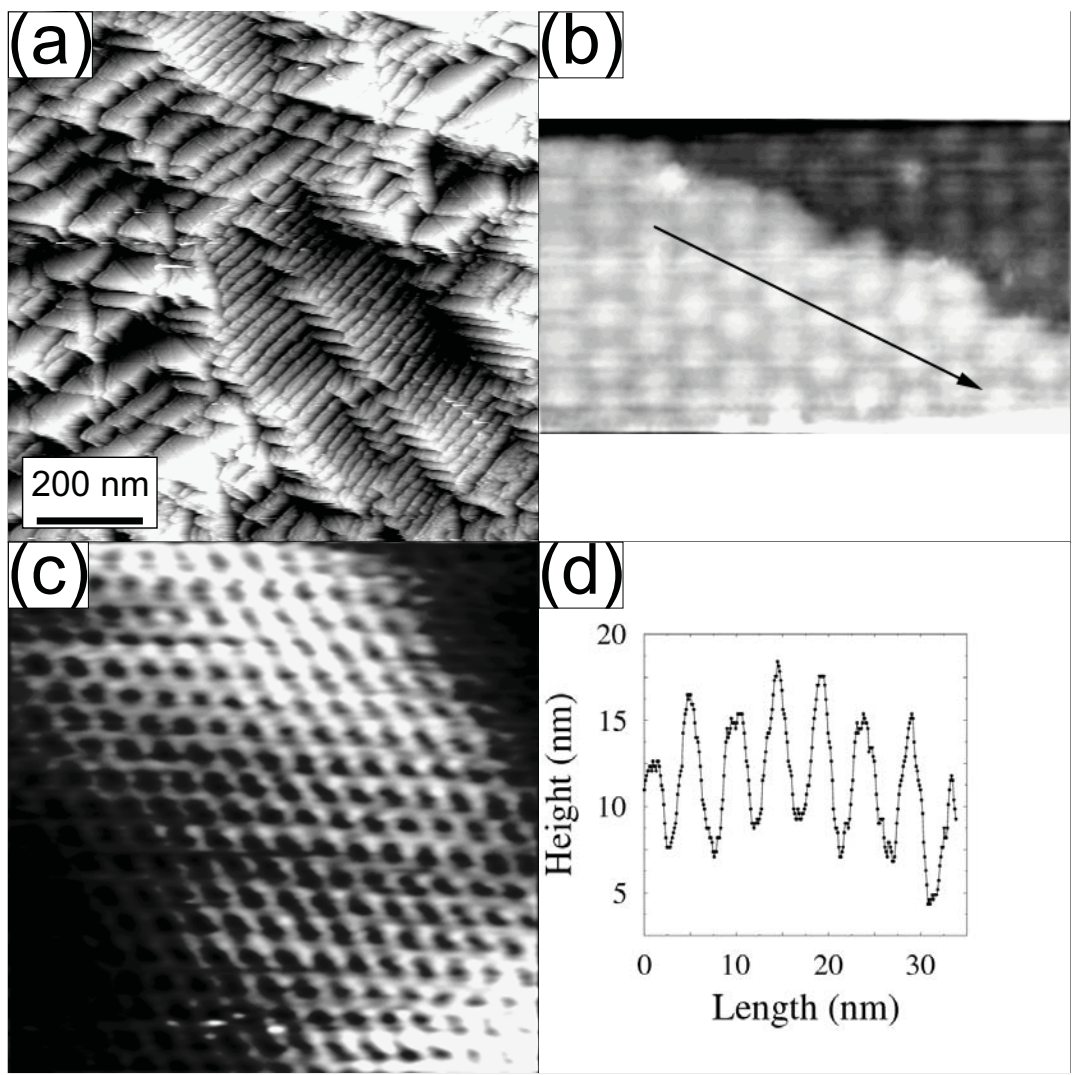

Fig 6. by Xu *et al*.